\documentclass[%
 reprint,
 amsmath,amssymb,
 aps,
]{revtex4-2}

\usepackage{graphicx}
\usepackage{dcolumn}
\usepackage{bm}
\usepackage[utf8]{inputenc}
\usepackage{ulem}

\newcommand*\diff{\mathop{}\!\mathrm{d}}

\usepackage{amsmath}
\usepackage{hyperref}
\usepackage{cleveref}
\usepackage{xparse}
\usepackage{physics}
\usepackage[caption=false]{subfig}
\usepackage{graphicx}
\usepackage{xcolor}

\usepackage{csquotes}

\graphicspath{{./}{figures/}}

\begin{document}

\preprint{APS/123-QED}

\title{Dynamic Phase Alignment in Inertial Alfv\'en Turbulence}

\author{Lucio M. Milanese}
\email{milanese@mit.edu}
\affiliation{Plasma Science and Fusion Center, Massachusetts Institute of Technology, Cambridge, MA 02139, USA}

\author{Maximilian Daschner}
\affiliation{Plasma Science and Fusion Center, Massachusetts Institute of Technology, Cambridge, MA 02139, USA}
\affiliation{ETH Zurich, CH-8093 Zürich, Switzerland}

\author{Nuno F. Loureiro}
\affiliation{Plasma Science and Fusion Center, Massachusetts Institute of Technology, Cambridge, MA 02139, USA}

\author{Stanislav Boldyrev}
\affiliation{Department of Physics, University of Wisconsin-Madison, Madison, WI 53706, USA}
\affiliation{Space Science Institute, Boulder, CO 80301, USA}

\date{\today}

\begin{abstract}
In weakly-collisional plasma environments with sufficiently low electron beta, Alfv\'enic turbulence transforms into inertial Alfv\'enic turbulence at scales below the electron skin-depth, $k_\perp d_e\gtrsim 1$. We argue that, in inertial Alfv\'enic turbulence, both energy and generalized kinetic helicity exhibit direct cascades. We demonstrate that the two cascades are compatible due to the existence of a strong scale-dependence of the phase alignment angle between velocity and magnetic field fluctuations, with the phase alignment angle scaling as $\cos\alpha_k\propto k_{\perp}^{-1}$. The kinetic and magnetic energy spectra scale as $\propto k_{\perp}^{-5/3}$ and $\propto k_{\perp}^{-11/3}$, respectively. As a result of the dual direct cascade, the generalized-helicity spectrum scales as $\propto k^{-5/3}_{\perp}$, implying progressive balancing of the turbulence as the cascade proceeds to smaller scales in the $k_{\perp} d_e \gg 1$ range. Turbulent eddies exhibit a phase-space anisotropy  $k_{\parallel} \propto k_{\perp}^{5/3}$, consistent with critically-balanced inertial Alfv\'en fluctuations. Our results may be applicable to a variety of geophysical, space, and astrophysical environments, including the Earth's magnetosheath and ionosphere, solar corona, non-relativistic pair plasmas, as well as to strongly rotating non-ionized fluids.    
  
\end{abstract}

\maketitle

\textit{Introduction.}~Many important turbulent plasma environments are characterized by a low ratio of the electron plasma pressure to magnetic energy density, that is, low $\beta_e$, in addition to weak collisionality. Examples are the ionosphere \cite{Goertz1979Magnetosphere-IonosphereCoupling,Kumari2014InertialAurora}, the Earth's magnetosheath \cite{Chen2017NatureMagnetosheath}, the solar corona \cite{Aschwanden2001TheCorona,Cranmer2009CoronalHoles} and some instances of the solar wind \cite{Smith2001DayDissipation,Chen2014Ion-scaleBeta}. Turbulence may play a role in structure formation, energy dissipation, magnetic reconnection, heat conduction, and other processes relevant for the dynamics and thermodynamics of such systems  \cite{Similon1990PlasmaTurbulence, Biskamp2003MagnetohydrodynamicTurbulence,Phan2018ElectronMagnetosheath,Vega2020Electron-onlyTurbulence,Stawarz2019PropertiesMagnetosheath,SharmaPyakurel2019TransitionTurbulence,Kozak2011StatisticalMagnetosheath, Wang2012SpatialSheet, Iwai2014CoronalObservations, Smith2001DayDissipation}. Despite vigorous investigation, the nature of turbulent fluctuations in low beta regimes remains incompletely understood and continues to attract considerable interest
\cite{Lucek2005TheMagnetosheath,Sahraoui2013ScalingTurbulence,Chen2017NatureMagnetosheath, Hughes2017Particle-in-cell, Passot2018GyrofluidTurbulence}. 

At scales below the electron skin depth in plasmas with sufficiently low $\beta_e$, the dominant low-frequency plasma modes are arguably nonlinear inertial Alfv\'en waves, whose turbulent cascade is governed by the existence of two ideal invariants: energy and generalized kinetic helicity. Turbulent dynamics in the presence of two invariants is poorly understood in both plasmas and non-ionized fluids \cite{Matthaeus2015IntermittencyPlasmas,Alexakis2018CascadesFlows}. It is possible that both invariants are subject to a forward (direct) cascade, or that one of them cascades forward and the other backward~\cite{Brissaud1973HelicityTurbulence,Chen2003IntermittencyHelicity,Matthaeus2015IntermittencyPlasmas,Podesta2010THEORYCROSS-HELICITY,Passot2018GyrofluidTurbulence,Alexakis2018CascadesFlows}.~When both quantities cascade forward, one can argue in favor of the cascade of either invariant setting the nonlinear eddy turn-over time \cite{Alexakis2018CascadesFlows}, greatly complicating the analysis and leading to different predictions and understanding of the underlying turbulent dynamics.

In this Letter, we propose that, in inertial Alfv\'en turbulence, both energy and (kinetic) helicity cascade forward, and it is the cascade of energy, rather than that of helicity, that determines the cascade time. We demonstrate that, rather remarkably, this is achieved via a strongly scale-dependent {\it phase alignment} between fluctuations of electric and magnetic potentials, which manages to suppress helicity while allowing the energy cascade to proceed unhindered. Our phenomenological model predicts the spectra of magnetic, kinetic, and helicity fluctuations in the inertial kinetic regime, shown here to be in good agreement with the results of numerical simulations.

More broadly, we conjecture that the phenomenon of scale-dependent phase alignment uncovered in this work may be the mechanism underpinning the joint forward cascade of two ideal invariants in other physical systems, including nonconducting fluids described by the Navier-Stokes equation \cite{Chen2003IntermittencyHelicity,Sahoo2017HelicityModels,Alexakis2018CascadesFlows,Pouquet2019HelicityReview}.

\textit{Model equations}.~
We consider a plasma permeated by a strong magnetic field, $B_0\vu*{z}$, such that the total field is $\vb{B} = B_0 \vu*{z} +  \vb{\delta B_{\perp}}$, with $\delta B_{\perp}/B_0\ll 1$. 
The evolutionary equations that we adopt are:
\begin{equation}
\pdv{}{t}  \laplacian_{\perp}\phi + \pb{\phi}{\laplacian_{\perp} \phi}  = \pb{\psi}{\laplacian_{\perp}\psi} + V_A \pdv{}{z} \laplacian_{\perp}\psi + f_{\phi} \label{Eq:phi}, 
\end{equation}
\begin{equation}
  \pdv{}{t}()  (1-d_e^2\laplacian_{\perp}) \psi + \pb{\phi}{(1-d_e^2\laplacian_{\perp})\psi}  =  V_A \pdv{\phi}{z} + f_{\psi}. \label{Eq:psi}
\end{equation}
Here, $\phi$ denotes the stream function, related to the $\textbf{E}\times \textbf{B}$ flow velocity by $\vb{v_{\perp}}=\vu*{z} \times {\bm\nabla}_{\perp} \phi$, and $\psi$ is the flux function, related to the perpendicular component of the magnetic field by $\delta \vb{B_{\perp}}/\sqrt{4 \pi \rho}=\vu*{z} \times {\bm\nabla}_{\perp} \psi$,
with $\rho$ the mass density. The Poisson bracket is defined as $\pb{A}{B}= \partial_x A \partial_y B - \partial_x B \partial_y A$, the Alfv\'en speed is $V_{A}=B_{0} / \sqrt{4 \pi \rho}$, and $f_{\phi}$ and $f_{\psi}$ are forcing terms to be described later. 
The only kinetic effect included in these equations is the electron inertia, characterized by $d_e$, the electron skin-depth \footnote{The dispersion relation for inertial Alfv\'en waves, Eq. (\ref{eq:eigenfunction}), implies that $\omega_l/|k_z| \gg v_{the}$ at scales $k_{\perp} \rho_e \ll 1$, where $\rho_e = m_e c v_{the}/e B_0$ is the electron Larmor radius. Since low-$\beta_e$ implies $d_e \gg \rho_e$, we find that electron Landau damping may thus be neglected at scales $d_e > k_{\perp}^{-1} \gg \rho_e$, where the resonant condition is not satisfied. This is unlike $\beta_e\sim 1$ plasmas, in which $d_e \sim \rho_e$ and where, therefore, electron Landau damping may be expected to play an important role at electron scales \cite{Sahraoui2009EvidenceGyroscale,Alexandrova2013SolarInstabilities}.}.

These equations describe low beta non-relativistic pair-plasmas~\cite{Loureiro2018TurbulencePlasmasc}, as well as electron-ion plasmas in the `ultralow' beta limit, $\beta_{e} \sim \beta_{i} \ll m_e/m_i$~\cite{Zocco2011}. The modes described by these equations are (as we show below) the inertial Alfv\'en modes. 
However, quite importantly, our equations are also pertinent to a wide range of other environments. 
When $k_{\perp}^2 d^2_e \gg 1+2/\beta_i$, our equations are structurally identical to Eqs.~(19) and (20) of Ref.~\cite{Chen2017NatureMagnetosheath} which were derived under the assumptions of $\beta_i \sim 1$ and $\beta_e \ll 1$. The dominant low-frequency modes there are inertial kinetic Alfv\'en waves ($\omega < k_{\perp} v_{thi}$, with $v_{thi}$ the ion thermal speed)  \cite{Chen2017NatureMagnetosheath,Boldyrev2019RoleTurbulence}. In addition, in the limit $k^2_{\perp} d^2_e \gg 1$, Eqs.~(\ref{Eq:phi}-\ref{Eq:psi}) are structurally equivalent to Eqs.~(25) and (26) of Ref.~\cite{Chen2017NatureMagnetosheath}, which describe inertial whistler waves ($\omega > k_{\perp} v_{thi}$) in reduced electron MHD. One can also demonstrate that, quite remarkably, in the limit $k_\perp^2 d_e^2\gg1$, our equations map onto the equations describing rapidly rotating non-ionized fluids \cite{Julien2007ReducedConstraints}. A short derivation of model Eqs.~(\ref{Eq:phi}-\ref{Eq:psi}) is presented in the Supplemental Material, where a summary of their regimes of applicability is also included.

 Eqs. (\ref{Eq:phi}-\ref{Eq:psi}) have two quadratic invariants: total energy, 
\begin{equation}\label{eq:energy_invariant}
    \mathcal{E}=\frac{1}{2} \int d V\left\{\left(\nabla_{\perp} \psi\right)^{2}+d_{e}^{2}\left(\nabla_{\perp}^{2} \psi\right)^{2}+\left(\nabla_{\perp} \phi\right)^{2}\right\},
\end{equation}
and generalized kinetic helicity,  
\begin{equation}\label{eq:kinetic helicity}
\mathcal{H}=\int d V\left\{\nabla_{\perp}^{2} \phi\left(1-d_{e}^{2} \nabla_{\perp}^{2}\right) \psi\right\},
\end{equation}
which reduces to cross-helicity at MHD scales ($k_{\perp} d_e \ll 1$).

The only linear mode supported by  these equations is the inertial Alfv\'en wave, with dispersion relation and eigenfunctions given by:
\begin{equation}
    \omega_{l}=\pm \frac{k_{z} V_{A}}{\sqrt{1+k_{\perp}^{2} d_{e}^{2}}}, \ \ \ \phi = \pm \sqrt{1+k_{\perp}^2 d_e^2}\, \psi. \label{eq:eigenfunction}
\end{equation}

\begin{figure*}[ht]
\centering 
\includegraphics[width=0.99\textwidth]{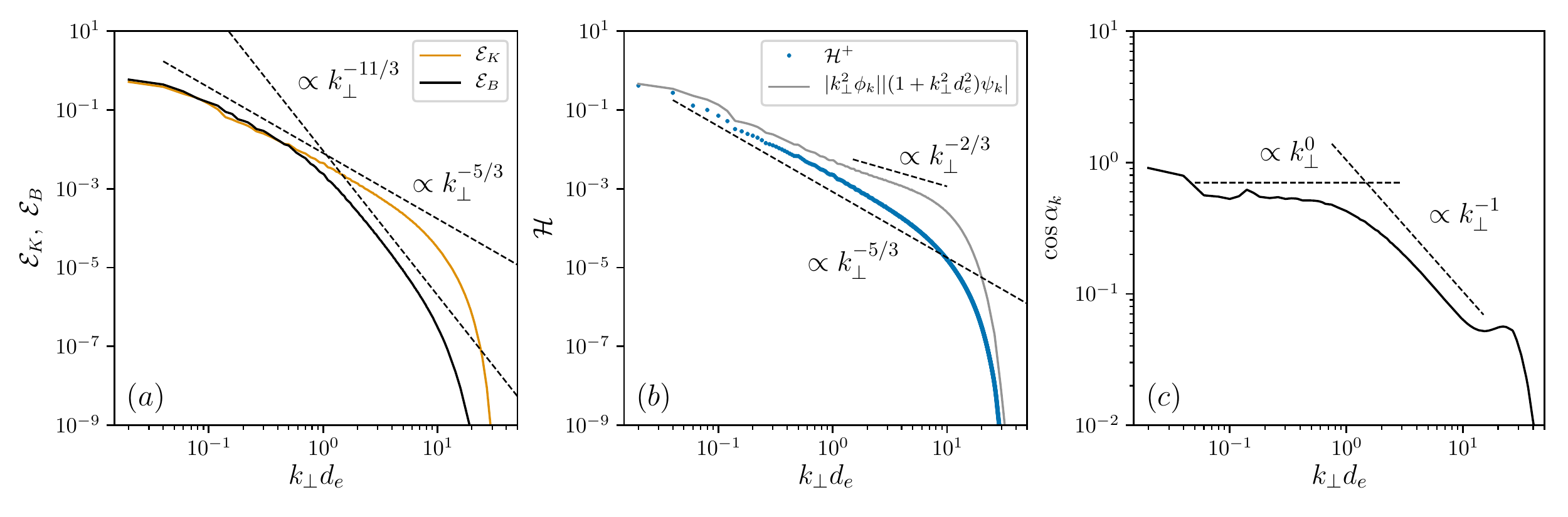}%
\caption{Simulation A1. (a) Spectra of magnetic and kinetic energy. (b) Spectra of kinetic helicity and of the product of the absolute value of the factors in the integrand of Eq.~(\ref{eq:kinetic helicity}). ~(c) Scale dependence of the average phase angle between fluctuations of electric and magnetic potential in Fourier space.}
\label{fig:plots_2048}%
\end{figure*}

\textit{Inertial Alfv\'en turbulence}.~The focus of our Letter is on turbulence in the kinetic range $k_\perp d_e \gg1$. In the opposite limit of $k_{\perp} d_e \ll 1$, Eqs.~(\ref{Eq:phi}-\ref{Eq:psi}) become the reduced MHD (RMHD) equations \citep{Kadomtsev1974NonlinearTokamak,Strauss1976,Schekochihin2009AstrophysicalPlasmas}, and thus results obtained for RMHD are expected to apply \citep{Perez2012OnTurbulence}.
Following Ref.~\cite{Loureiro2018TurbulencePlasmasc}, the energy flux at scales $k_{\perp} d_e > 1$ is expected to be $\varepsilon \sim k_{\perp}^{2} \phi_{\lambda}^{2} / \tau_{\lambda}$, where $\tau_{\lambda}$ is the eddy turnover time at the scale $\lambda \sim k_{\perp}^{-1}$, and $\tau_{\lambda}= 1 / \omega_{n l} \sim 1 /\left(k_{\perp}^{2} \phi_{\lambda}\right)$. This yields $\phi_{\lambda} \sim \varepsilon^{1 / 3} k_{\perp}^{-4 / 3}$, leading to the scaling of the spectrum of perpendicular kinetic energy $\mathcal{E}_{K}\left(k_{\perp}\right) d k_{\perp} \sim \varepsilon^{2 / 3} k_{\perp}^{-5 / 3} d k_{\perp}$. For $k^2_{\perp} d^2_e \gg 1$, equipartition between parallel and perpendicular kinetic energies, i.e., between the second and third terms in Eq.~(\ref{eq:energy_invariant}), results in $\psi_{\lambda} \sim k_{\perp}^{-7/3}$ from which follows the magnetic energy spectrum $\mathcal{E}_{B}\left(k_{\perp}\right) d k_{\perp} \sim \varepsilon^{2 / 3} k_{\perp}^{-11 / 3} d k_{\perp}$. 
Finally, postulating critical balance of the fluctuations in this range (i.e., that the characteristic linear and nonlinear frequencies of the system approximately balance at each scale \cite{Goldreich1995TowardTurbulence, Schekochihin2009AstrophysicalPlasmas}), yields
\begin{equation}\label{eq:scale_dependent_anisotropy}
    k_{\|} \sim \varepsilon^{1 / 3} d_{e} V_{A}^{-1} k_{\perp}^{5 / 3}.
\end{equation}

Using the above scalings for $\phi_\lambda$ and $\psi_\lambda$, we would predict the helicity spectrum to scale as  $\mathcal{H}\left(k_{\perp}\right) d k_{\perp} \sim k_{\perp}^{-2 / 3} d k_{\perp}$ in the kinetic range. However, as discussed in Ref. \cite[]{Loureiro2018TurbulencePlasmasc}, in this case the helicity flux cannot be constant; rather it should \textit{increase} at small scales, leading to a contradiction. If, on the other hand, we assume that the scaling of the fields should be determined by a direct helicity cascade, we would formally conclude that the energy cannot cascade toward small scales at $k_{\perp} d_e>1$. 
This contradiction can be solved if, as conjectured in Ref. \cite{Loureiro2018TurbulencePlasmasc}, the helicity flux at scales $k_\perp d_e\gg1$ is written as 
\begin{equation}\label{eq:first_helicity_scaling}
    \left(k_{\perp}^{2} \phi_{\lambda}\right)\left(d_{e}^{2} k_{\perp}^{2} \psi_{\lambda}\right) R_{\lambda} / \tau_{\lambda} \sim \varepsilon_{H},
\end{equation}
where $R_{\lambda}$ is a scale-dependent cancellation factor.
Requiring that the flux of kinetic helicity be constant in the kinetic range and enforcing consistency between energy and helicity fluxes leads to  
\begin{equation}
R_{\lambda} \sim \varepsilon_{H} \left(k_{\perp}^{2} \phi_{\lambda}\right)^{-2}\left(d_{e}^{2} k_{\perp}^{2} \psi_{\lambda}\right)^{-1} \sim  \varepsilon_{H} (k_{\perp} d_e)^{-1}.
\end{equation}

When the cancellation factor is present, the simultaneous direct cascades of \textit{both} energy and helicity become possible, and we arrive at a qualitatively different prediction for the helicity spectrum, $\mathcal{H}\left(k_{\perp}\right) d k_{\perp} \sim k_{\perp}^{-5 / 3} d k_{\perp}$. In what follows we demonstrate that the cancellation factor is a manifestation of a new phenomenon that we call \textit{``dynamic phase alignment''}: an increasing correlation between the \textit{phases} of the fluctuating magnetic and velocity fields as the cascade progresses towards smaller scales.

\textit{Numerical setup.} We now report on direct numerical simulations carried out to test these theoretical predictions. We integrate Eqs. (\ref{Eq:phi}-\ref{Eq:psi}) with the code \texttt{Viriato} \citep{Loureiro2016} on a triply periodic domain using a grid of $N_\perp^2\times N_\parallel$ points. 
Hyper-dissipation terms of the form $\nu_H \nabla^6_{\perp}$ are included on the right-hand side of both equations, with $\nu_H$ set to remove energy at the grid scale. Energy is injected via delta-correlated forcing terms of the form
\begin{equation}\label{eq:forcing_terms}
\begin{split}
 f_{\phi,\psi}  = C_{\phi,\psi} \alpha_{\pm}\delta  (k_x - k_{x0}) \delta (k_y - k_{y0} ) \cos(k_{z0} z),
 \end{split}
\end{equation}
where $C_{\phi}$ and $C_{\psi}$ are randomly chosen complex numbers determining the phase of the mode being excited ($C_{\phi} \neq C_{\psi}$, and $|C_{\phi,\psi}|=1$), and $\alpha_{\pm} > 0$ are numerical coefficients determining the strength of the drive, their subscript relating to positive and negative (generalized) kinetic helicity injection, as discussed below. 
The mode numbers $k_{x0}$, $k_{y0}$ and $k_{z0}$ are randomly chosen from a predetermined range and at every time step they are the same for both $f_{\phi}$ and $f_{\psi}$.

From Eqs.~(\ref{eq:kinetic helicity}, \ref{eq:forcing_terms}), one can show that the kinetic helicity injected at any time step by the forcing terms is given by 
\begin{equation}
    \mathcal{H}_{inj}^{\pm} \propto k^2_{\perp 0}(1 + k^2_{\perp 0} d_e^2) \alpha_{\pm}^{2} \Re[C_{\phi}C_{\psi}],
\end{equation}
where $k_{\perp0}^2 = k_{x0}^2 + k_{y0} ^ 2$. When $\Re[C_{\phi}C_{\psi}] > 0$, i.e., when the phase between $f_{\phi}$ and $f_{\psi}$ is such that positive helicity is injected at a particular time step, the $\alpha_{+}$ coefficient is used in the forcing terms.
~When $\Re[C_{\phi}C_{\psi}] < 0$, the coefficient  $\alpha_{-}$ is used instead. We define the ratio of positive to negative kinetic helicity injection as $ \mathcal{R}_{\mathcal{H}} \equiv \mathcal{H}_{inj}^{+}/\mathcal{H}_{inj}^{-}$. The ratio of the coefficients is set as  $\alpha_{+}/\alpha_{-} = \sqrt{\mathcal{R}_{\mathcal{H}}}$.

\begin{figure*}
\centering 
\includegraphics[width=0.99\textwidth]{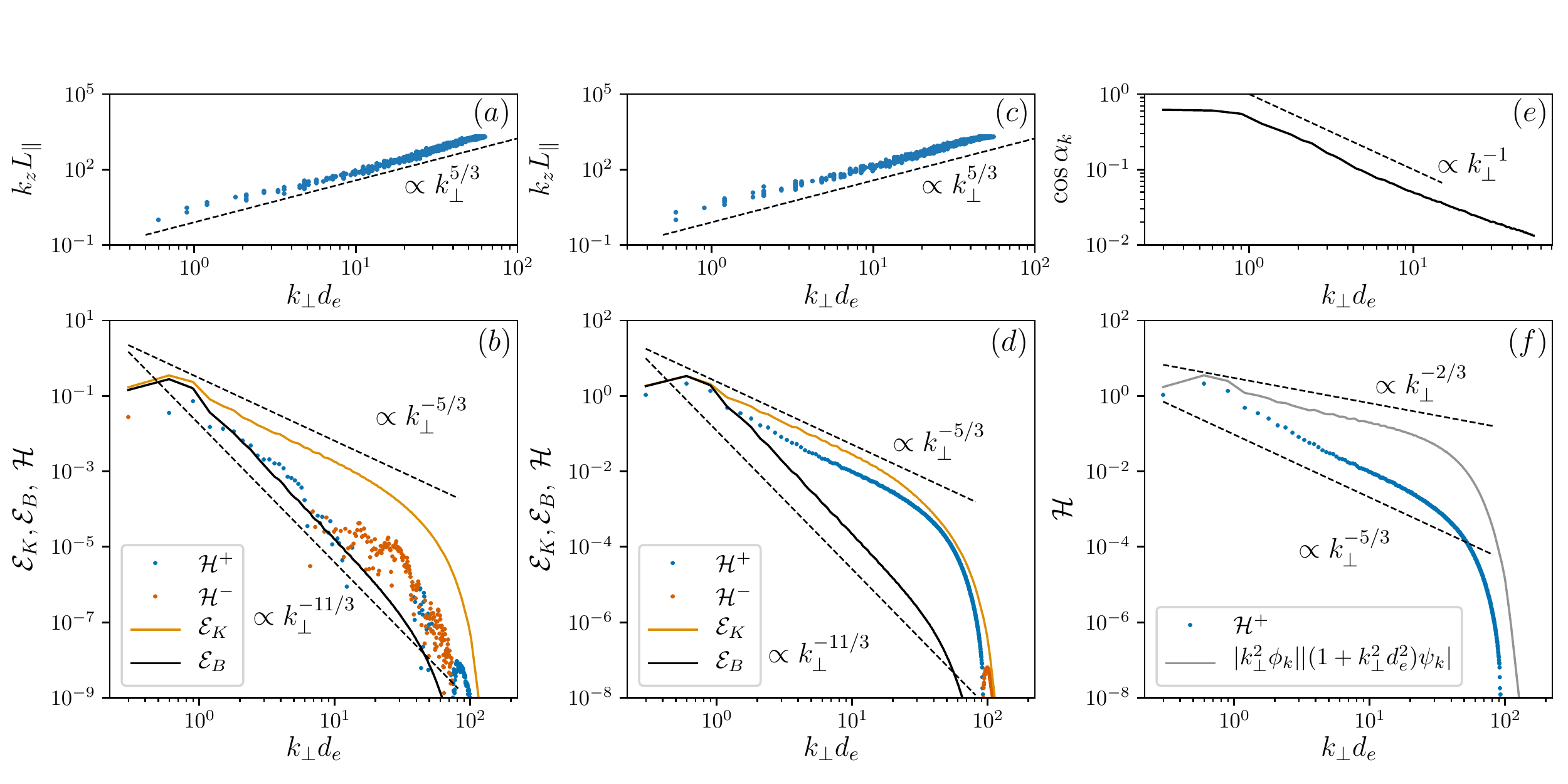}
\caption{Eddy anisotropy scaling for simulations B1 and B2 (subplots (a) and (c), respectively). Spectra of kinetic helicity, and of kinetic and magnetic energy for simulations B1 and B2 (subplots (b) and (d), respectively). Different colors are used to represent the presence of net positive or negative helicity in perpendicular wavenumber shells. Subplot (f) presents the spectra of kinetic helicity and of the product of the absolute value of the factors in the integrand of Eq.~(\ref{eq:kinetic helicity}) in simulation B2. The corresponding  average value of the cosine of the phase angle (Eq.~(\ref{eq:cos_alpha})) as a function of scale is shown in subplot (e).}
\label{plot:comparison_plots}
\end{figure*}

Table \ref{tab:simulation_parameters} summarizes key parameters of the simulations performed.  In all cases, energy is injected at the largest scales, where $k_{\perp} d_e < 1$. In runs A1 and B2 net positive kinetic helicity is injected by the forcing terms ($\mathcal{R}_{\mathcal{H}} = 10$ and $\mathcal{R}_{\mathcal{H}} = 30$, respectively), while in run B1 no \textit{net} kinetic helicity is injected ($\mathcal{R}_{\mathcal{H}} = 1$). Run A1 aims at capturing the dynamics in both the RMHD and kinetic range, and providing insight into how the transition between the two regimes occurs. Simulations of type B aim at capturing in more detail the turbulent dynamics in the kinetic range.
\begin{table}[h]
\begin{ruledtabular}
\begin{tabular}{ccccc}
ID & $N_{\perp}$ & $N_{\parallel}$ &   $(k_{\perp} d_e)_{min}$  & $\mathcal{R}_{\mathcal{H}}$\\
\colrule
A1 & 2048 & 2048 & 0.02 & 10 \\
\hline
B1 & 768 & 4096 & 0.3 & 1    \\ 
B2 & 768 & 4096 & 0.3 & 30  \\
\end{tabular}
\caption{\label{tab:simulation_parameters}%
Summary of key simulation parameters.}
\end{ruledtabular}
\end{table}

\textit{Energy spectra.}
 Figs.~\ref{fig:plots_2048}a and \ref{fig:plots_2048}b show the energy and (generalized kinetic) helicity spectra (obtained from time-averaged data after steady state is reached) for simulation A1. 
 The magnetic energy spectrum is seen to smoothly transition from $\sim k_{\perp}^{-5/3}$ to $\sim k_{\perp}^{-11/3}$ at $k_{\perp} d_e \approx 1$, whereas the kinetic energy scales as $\sim k_\perp^{-5/3}$ throughout the inertial range, as does the helicity spectrum. These observations are in good agreement with the theoretical predictions, and offer an immediate confirmation of the existence of the scale-dependent cancellation factor $R_\lambda\sim 1/k_\perp$ at scales $k_\perp d_e > 1$.
 
Runs of type B confirm the kinetic range results over a larger scale range; see Figs.~\ref{plot:comparison_plots}b and \ref{plot:comparison_plots}d. 
The energy spectra are not significantly affected by the ratio of positive to negative helicity injected in the system. 
When no net helicity is injected in the system, the spectrum of helicity is not well defined (Fig.~\ref{plot:comparison_plots}b). One can observe that the sign of kinetic helicity is different at different perpendicular wavenumbers $k_{\perp}$ in the inertial range, and its value is zero when spatially averaged over the entire simulation domain and time averaged over the steady state. When instead net helicity is injected in the system, a well-defined spectrum is observed, exhibiting a scaling $\sim k_{\perp}^{-5 / 3}$ (Fig.~\ref{plot:comparison_plots}d), as in simulation A1.

To characterize eddy anisotropy, we consider that the parallel wavenumber of a fluctuating field $\phi$ at perpendicular scale $k_{\perp}$ may be approximated as~\cite{Cho2004TheTurbulence} 
\begin{equation}\label{k_||}
    k_{\parallel} \approx\left(\frac{\left\langle\left|\mathbf {B}_{0} \partial_z \phi_ {k_{\perp} } + \delta \mathbf{B}_{k_{\perp}} \cdot {\bm\nabla} \phi_ {k_{\perp}}\right|^{2}\right\rangle}{\left\langle B_{k_{\perp}}^{2}\right\rangle\left\langle \phi_ {k_{\perp}}^{2}\right\rangle}\right)^{1 / 2},
\end{equation}
where $\left\langle ... \right\rangle$ denotes spatial averaging. 
In the kinetic range, electromagnetic fluctuations are small because electron inertia ($k_{\perp}^2 d_e^2 \partial_t \psi$) dominates over the inductive part of the electric field ($\partial_t \psi$) in Eq.~(\ref{Eq:psi}).~ Therefore, turbulence in this regime is essentially electrostatic, i.e., $\mathbf{B}_{0} \partial_z \phi_ {k_{\perp} } \gg \delta \mathbf{B}_{k_{\perp}} \cdot \bm \nabla \phi_ {k_{\perp}}$, and thus $k_\| \approx k_z$. 
The scatter plots in Figs.~\ref{plot:comparison_plots}a and \ref{plot:comparison_plots}c show, for each value of $k_z$, the corresponding value of $k_\perp$ at which the energy of the $\phi$ fluctuations is largest. The data exhibit the scaling $k_{\parallel} \propto k_{\perp}^{5/3}$, in agreement with Eq. (\ref{eq:scale_dependent_anisotropy}), confirming that the inertial Alfv\'en cascade is critically balanced. 

\begin{figure}
\centering 
\includegraphics[width=0.99\columnwidth]{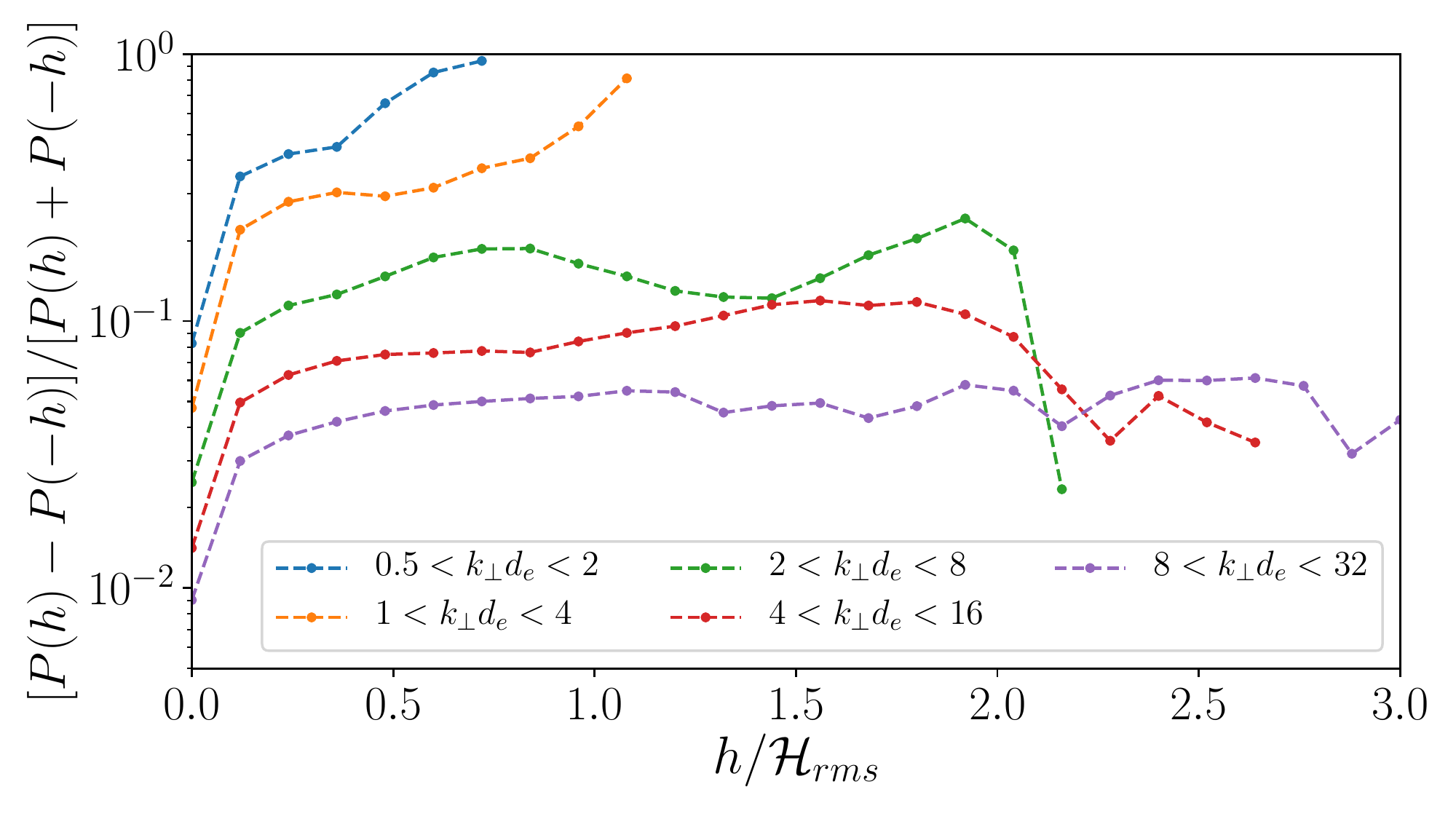}
\caption{Relative difference between the PDF of positive and negative helicity density obtained from band-pass filtered data from simulation B2.}
\label{fig:helicity_pdf_comparison}
\end{figure}

\textit{Kinetic helicity spectrum and dynamic phase alignment.} 
The net kinetic helicity at each wavenumber is a function of the absolute value of the Fourier coefficients $|\phi_{k}|$ and $|\psi_{k}|$, and of the phase angle between them, $\alpha_k$. For a given $k_\perp$, the average value of $\alpha_k$ is given by
\begin{equation}
\label{eq:cos_alpha}
\cos\alpha_k = \frac{1}{2} \frac{\langle  k_{\perp}^{2} \phi_{k \perp}(1+d_e^2k_{\perp}^{2})\psi_{k \perp}^{*} + c. c. \rangle}{\langle|k^2_{\perp} \phi_{k \perp}||(1+d_e^2k_{\perp}^{2})\psi_{k \perp}| \rangle},
\end{equation} where the numerator represents net kinetic helicity at a given perpendicular wavenumber.~In the $k_{\perp} d_e < 1$ range, the spectra of  $|k_{\perp}^2\phi_{k \perp}||\psi_{k \perp}|$ and of generalized kinetic helicity (which turns into cross helicity at such scales) are both expected to exhibit the scaling of the MHD energy spectrum, and thus $\cos\alpha_k$ should not depend strongly on scale. At scales $k_{\perp} d_e > 1$, however, the spectrum of $|k_{\perp}^2\phi_{k \perp}||k_{\perp}^2 d^2_e\psi_{k \perp}|$ is expected to scale as $k_{\perp}^{-2/3}$,  while we predict, and observe, kinetic helicity to scale as $k_{\perp}^{-5/3}$. We thus expect  $\cos\alpha_k \propto k_{\perp}^{-1}$. This is confirmed in Fig. \ref{fig:plots_2048}c: in the RMHD range,  $\cos\alpha_k $ does not vary strongly as a function of scale. After a smooth transition at $k_{\perp} d_e \approx 1$, the scaling of $\cos\alpha_k $ asymptotes to $\propto k_{\perp}^{-1}$ for $k_{\perp} d_e > 1$. Fig.~\ref{plot:comparison_plots}e confirms the scaling $\cos\alpha_k \propto k_{\perp}^{-1}$ in the kinetic range. 

When $\cos\alpha_k=0$, kinetic helicity is zero and the system is in a perfectly balanced state. The scaling $\cos\alpha_k \propto k_{\perp}^{-1}$ therefore implies that the turbulence becomes progressively more balanced as the cascade proceeds deeper in the kinetic range.
This statement is corroborated by results from simulation B2 shown in Fig.~\ref{fig:helicity_pdf_comparison}, in which we present the scale-dependence of asymmetries in the probability density function (PDF) of generalized kinetic helicity density ($h$). We plot, at different perpendicular scales,  the relative difference between the PDF of positive and negative helicity density, i.e., $P(h)$ and $P(-h)$, for $h \in [0, 3 \ \mathcal{H}_{rms}]$, where $\mathcal{H}_{rms}$ is the root mean square value of helicity density obtained from unfiltered data.~As the selection of scales included in the band-pass filter moves towards larger values of $k_{\perp}$, the relative difference between $P(h)$ and $P(-h)$ becomes smaller, showing that the PDF is progressively more symmetric and thus more balanced (a perfectly symmetric PDF implies that the turbulence is balanced, as $\mathcal{H} = \int P(h) h \diff h = 0$).

\textit{Conclusions.} In this Letter, we showed that, in inertial Alfv\'en turbulence, both energy and generalized kinetic helicity cascade forward, with the cascade of energy determining the nonlinear eddy turn-over time. Helicity is found to scale as $\mathcal{H} \propto k_{\perp}^{-5/3}$ in the kinetic range, a result that is underpinned by a scale-dependent alignment angle, $\cos\alpha_k \propto k_{\perp}^{-1}$, between the Fourier phases of magnetic and velocity fields. Consequently, turbulence becomes progressively more balanced as the cascade proceeds deeper into the kinetic range. 

The results presented in this Letter may be valuable for interpreting the direct measurements of low-beta turbulence in space plasmas \cite{Aschwanden2001TheCorona, Smith2001DayDissipation, Cranmer2009CoronalHoles, Chen2014Ion-scaleBeta}, as well as for other astrophysical and geophysical turbulent systems where dual energy and kinetic helicity cascades are possible (e.g., sub-relativistic pair plasma \cite{Loureiro2018TurbulencePlasmasc}, whose experimental realization is upcoming \cite{ Stenson2018LosslessTrap, Horn-Stanja2020InjectionTrapb}, ionospheric \cite{Goertz1979Magnetosphere-IonosphereCoupling,Kumari2014InertialAurora} and magnetospheric plasmas \cite{Chen2017NatureMagnetosheath}, and strongly rotating nonconducting fluids \cite{Nazarenko2011CriticalConjecture}).~Another context where our findings may be pertinent is Navier-Stokes (NS) turbulence. Simulations reveal a $k^{-5/3}$ scaling of kinetic helicity and a scale-dependent progressive balancing of turbulence (restoration of mirror symmetry) \cite{Chen2003IntermittencyHelicity,Sahoo2017HelicityModels, Pouquet2019HelicityReview} whose underlying dynamics is not fully understood. We conjecture that the novel mechanism of dynamic phase alignment uncovered in this work may also be at play in NS turbulence, and account for those results.
While the details of the nonlinear interactions in plasma and NS turbulence are different, our conjecture is based on commonalities between particular aspects of the joint forward cascade of energy and (generalized) kinetic helicity. In particular, in both systems, a `na\"ive' estimate of the spectral scaling of helicity, without the inclusion of a scale-dependent phase alignment factor, would yield a scaling $\sim k^{-2/3}$, which, if realized, would prevent energy from cascading forward. In both systems a scaling $\mathcal{H} \sim k^{-5/3}$ is instead observed \cite{Chen2003IntermittencyHelicity, Pouquet2019HelicityReview}, which may be underpinned, in the case of NS turbulence, by a scale-dependent alignment between the phases of velocity and vorticity fluctuations.

Work supported by DoE grant No.~DE-FG02-91ER54109 (L.M.M. and N.F.L.), NSF CAREER award No. 1654168 (N.F.L.), the Prof. Amar G. Bose Research Fellows Program at MIT (L.M.M. and N.F.L.), and NSF grant No. PHY-1707272,
NASA grant No. 80NSSC18K0646, and DOE grant No.
DESC0018266 (S.B.). This research used resources  of the facilities of the Massachusetts Green
High-Performance Computing Center (MGHPCC) and of the National Energy Research Scientific Computing Center (NERSC), a U.S. Department of Energy Office of Science User Facility operated under Contract No. DE-AC02-05CH11231.

\bibliography{references_PRL2}

\begin{thebibliography}{44}%
\makeatletter
\providecommand \@ifxundefined [1]{%
 \@ifx{#1\undefined}
}%
\providecommand \@ifnum [1]{%
 \ifnum #1\expandafter \@firstoftwo
 \else \expandafter \@secondoftwo
 \fi
}%
\providecommand \@ifx [1]{%
 \ifx #1\expandafter \@firstoftwo
 \else \expandafter \@secondoftwo
 \fi
}%
\providecommand \natexlab [1]{#1}%
\providecommand \enquote  [1]{``#1''}%
\providecommand \bibnamefont  [1]{#1}%
\providecommand \bibfnamefont [1]{#1}%
\providecommand \citenamefont [1]{#1}%
\providecommand \href@noop [0]{\@secondoftwo}%
\providecommand \href [0]{\begingroup \@sanitize@url \@href}%
\providecommand \@href[1]{\@@startlink{#1}\@@href}%
\providecommand \@@href[1]{\endgroup#1\@@endlink}%
\providecommand \@sanitize@url [0]{\catcode `\\12\catcode `\$12\catcode
  `\&12\catcode `\#12\catcode `\^12\catcode `\_12\catcode `\%12\relax}%
\providecommand \@@startlink[1]{}%
\providecommand \@@endlink[0]{}%
\providecommand \url  [0]{\begingroup\@sanitize@url \@url }%
\providecommand \@url [1]{\endgroup\@href {#1}{\urlprefix }}%
\providecommand \urlprefix  [0]{URL }%
\providecommand \Eprint [0]{\href }%
\providecommand \doibase [0]{https://doi.org/}%
\providecommand \selectlanguage [0]{\@gobble}%
\providecommand \bibinfo  [0]{\@secondoftwo}%
\providecommand \bibfield  [0]{\@secondoftwo}%
\providecommand \translation [1]{[#1]}%
\providecommand \BibitemOpen [0]{}%
\providecommand \bibitemStop [0]{}%
\providecommand \bibitemNoStop [0]{.\EOS\space}%
\providecommand \EOS [0]{\spacefactor3000\relax}%
\providecommand \BibitemShut  [1]{\csname bibitem#1\endcsname}%
\let\auto@bib@innerbib\@empty
\bibitem [{\citenamefont {Goertz}\ and\ \citenamefont
  {Boswell}(1979)}]{Goertz1979Magnetosphere-IonosphereCoupling}%
  \BibitemOpen
  \bibfield  {author} {\bibinfo {author} {\bibfnamefont {C.~K.}\ \bibnamefont
  {Goertz}}\ and\ \bibinfo {author} {\bibfnamefont {B.~W.}\ \bibnamefont
  {Boswell}},\ }\bibfield  {title} {\bibinfo {title} {{Magnetosphere-Ionosphere
  coupling}},\ }\href {https://doi.org/10.1029/JA084iA12p07239} {\bibfield
  {journal} {\bibinfo  {journal} {Journal of Geophysical Research}\ }\textbf
  {\bibinfo {volume} {84}},\ \bibinfo {pages} {7239} (\bibinfo {year}
  {1979})}\BibitemShut {NoStop}%
\bibitem [{\citenamefont {Kumari}\ \emph {et~al.}(2014)\citenamefont {Kumari},
  \citenamefont {Sharma},\ and\ \citenamefont
  {Yadav}}]{Kumari2014InertialAurora}%
  \BibitemOpen
  \bibfield  {author} {\bibinfo {author} {\bibfnamefont {A.}~\bibnamefont
  {Kumari}}, \bibinfo {author} {\bibfnamefont {R.~P.}\ \bibnamefont {Sharma}},\
  and\ \bibinfo {author} {\bibfnamefont {N.}~\bibnamefont {Yadav}},\ }\bibfield
   {title} {\bibinfo {title} {{Inertial Alfven wave induced turbulent spectra
  in aurora}},\ }\href {https://doi.org/10.1007/s10509-014-1828-8} {\bibfield
  {journal} {\bibinfo  {journal} {Astrophysics and Space Science}\ }\textbf
  {\bibinfo {volume} {351}},\ \bibinfo {pages} {81} (\bibinfo {year}
  {2014})}\BibitemShut {NoStop}%
\bibitem [{\citenamefont {Chen}\ and\ \citenamefont
  {Boldyrev}(2017)}]{Chen2017NatureMagnetosheath}%
  \BibitemOpen
  \bibfield  {author} {\bibinfo {author} {\bibfnamefont {C.~H.~K.}\
  \bibnamefont {Chen}}\ and\ \bibinfo {author} {\bibfnamefont {S.}~\bibnamefont
  {Boldyrev}},\ }\bibfield  {title} {\bibinfo {title} {{Nature of Kinetic Scale
  Turbulence in the Earth's Magnetosheath}},\ }\href
  {https://doi.org/10.3847/1538-4357/aa74e0} {\bibfield  {journal} {\bibinfo
  {journal} {The Astrophysical Journal}\ }\textbf {\bibinfo {volume} {842}},\
  \bibinfo {pages} {122} (\bibinfo {year} {2017})}\BibitemShut {NoStop}%
\bibitem [{\citenamefont {Aschwanden}\ \emph {et~al.}(2001)\citenamefont
  {Aschwanden}, \citenamefont {Poland},\ and\ \citenamefont
  {Rabin}}]{Aschwanden2001TheCorona}%
  \BibitemOpen
  \bibfield  {author} {\bibinfo {author} {\bibfnamefont {M.~J.}\ \bibnamefont
  {Aschwanden}}, \bibinfo {author} {\bibfnamefont {A.~I.}\ \bibnamefont
  {Poland}},\ and\ \bibinfo {author} {\bibfnamefont {D.~M.}\ \bibnamefont
  {Rabin}},\ }\bibfield  {title} {\bibinfo {title} {{The new solar corona}},\
  }\href {https://doi.org/10.1146/annurev.astro.39.1.175} {\bibfield  {journal}
  {\bibinfo  {journal} {Annual Review of Astronomy and Astrophysics}\ }\textbf
  {\bibinfo {volume} {39}},\ \bibinfo {pages} {175} (\bibinfo {year}
  {2001})}\BibitemShut {NoStop}%
\bibitem [{\citenamefont {Cranmer}(2009)}]{Cranmer2009CoronalHoles}%
  \BibitemOpen
  \bibfield  {author} {\bibinfo {author} {\bibfnamefont {S.~R.}\ \bibnamefont
  {Cranmer}},\ }\bibfield  {title} {\bibinfo {title} {{Coronal holes}},\ }\href
  {https://doi.org/10.12942/lrsp-2009-3} {\bibfield  {journal} {\bibinfo
  {journal} {Living Reviews in Solar Physics}\ }\textbf {\bibinfo {volume}
  {6}},\ \bibinfo {pages} {3} (\bibinfo {year} {2009})}\BibitemShut {NoStop}%
\bibitem [{\citenamefont {Smith}\ \emph {et~al.}(2001)\citenamefont {Smith},
  \citenamefont {Mullan}, \citenamefont {Ness}, \citenamefont {Skoug},\ and\
  \citenamefont {Steinberg}}]{Smith2001DayDissipation}%
  \BibitemOpen
  \bibfield  {author} {\bibinfo {author} {\bibfnamefont {C.~W.}\ \bibnamefont
  {Smith}}, \bibinfo {author} {\bibfnamefont {D.~J.}\ \bibnamefont {Mullan}},
  \bibinfo {author} {\bibfnamefont {N.~F.}\ \bibnamefont {Ness}}, \bibinfo
  {author} {\bibfnamefont {R.~M.}\ \bibnamefont {Skoug}},\ and\ \bibinfo
  {author} {\bibfnamefont {J.}~\bibnamefont {Steinberg}},\ }\bibfield  {title}
  {\bibinfo {title} {{Day the solar wind almost disappeared: Magnetic field
  fluctuations, wave refraction and dissipation}},\ }\href
  {https://doi.org/10.1029/2001ja000022} {\bibfield  {journal} {\bibinfo
  {journal} {Journal of Geophysical Research: Space Physics}\ }\textbf
  {\bibinfo {volume} {106}},\ \bibinfo {pages} {18625} (\bibinfo {year}
  {2001})}\BibitemShut {NoStop}%
\bibitem [{\citenamefont {Chen}\ \emph {et~al.}(2014)\citenamefont {Chen},
  \citenamefont {Leung}, \citenamefont {Boldyrev}, \citenamefont {Maruca},\
  and\ \citenamefont {Bale}}]{Chen2014Ion-scaleBeta}%
  \BibitemOpen
  \bibfield  {author} {\bibinfo {author} {\bibfnamefont {C.~H.}\ \bibnamefont
  {Chen}}, \bibinfo {author} {\bibfnamefont {L.}~\bibnamefont {Leung}},
  \bibinfo {author} {\bibfnamefont {S.}~\bibnamefont {Boldyrev}}, \bibinfo
  {author} {\bibfnamefont {B.~A.}\ \bibnamefont {Maruca}},\ and\ \bibinfo
  {author} {\bibfnamefont {S.~D.}\ \bibnamefont {Bale}},\ }\bibfield  {title}
  {\bibinfo {title} {{Ion-scale spectral break of solar wind turbulence at high
  and low beta}},\ }\href {https://doi.org/10.1002/2014GL062009} {\bibfield
  {journal} {\bibinfo  {journal} {Geophysical Research Letters}\ }\textbf
  {\bibinfo {volume} {41}},\ \bibinfo {pages} {8081} (\bibinfo {year}
  {2014})}\BibitemShut {NoStop}%
\bibitem [{\citenamefont {Similon}\ and\ \citenamefont
  {Sudan}(1990)}]{Similon1990PlasmaTurbulence}%
  \BibitemOpen
  \bibfield  {author} {\bibinfo {author} {\bibfnamefont {P.~L.}\ \bibnamefont
  {Similon}}\ and\ \bibinfo {author} {\bibfnamefont {R.~N.}\ \bibnamefont
  {Sudan}},\ }\bibfield  {title} {\bibinfo {title} {{Plasma Turbulence}},\
  }\href {https://doi.org/10.1146/annurev.fl.22.010190.001533} {\bibfield
  {journal} {\bibinfo  {journal} {Annual Review of Fluid Mechanics}\ }\textbf
  {\bibinfo {volume} {22}},\ \bibinfo {pages} {317} (\bibinfo {year}
  {1990})}\BibitemShut {NoStop}%
\bibitem [{\citenamefont
  {Biskamp}(2003)}]{Biskamp2003MagnetohydrodynamicTurbulence}%
  \BibitemOpen
  \bibfield  {author} {\bibinfo {author} {\bibfnamefont {D.}~\bibnamefont
  {Biskamp}},\ }\href {https://doi.org/10.1017/cbo9780511535222} {\emph
  {\bibinfo {title} {Magnetohydrodynamic Turbulence}}}\ (\bibinfo  {publisher}
  {Cambridge University Press},\ \bibinfo {year} {2003})\BibitemShut {NoStop}%
\bibitem [{\citenamefont {Phan}\ \emph {et~al.}(2018)\citenamefont {Phan},
  \citenamefont {Eastwood}, \citenamefont {Shay}, \citenamefont {Drake},
  \citenamefont {Sonnerup}, \citenamefont {Fujimoto}, \citenamefont {Cassak},
  \citenamefont {{\O}ieroset}, \citenamefont {Burch}, \citenamefont {Torbert},
  \citenamefont {Rager}, \citenamefont {Dorelli}, \citenamefont {Gershman},
  \citenamefont {Pollock}, \citenamefont {Pyakurel}, \citenamefont {Haggerty},
  \citenamefont {Khotyaintsev}, \citenamefont {Lavraud}, \citenamefont {Saito},
  \citenamefont {Oka}, \citenamefont {Ergun}, \citenamefont {Retino},
  \citenamefont {Le~Contel}, \citenamefont {Argall}, \citenamefont {Giles},
  \citenamefont {Moore}, \citenamefont {Wilder}, \citenamefont {Strangeway},
  \citenamefont {Russell}, \citenamefont {Lindqvist},\ and\ \citenamefont
  {Magnes}}]{Phan2018ElectronMagnetosheath}%
  \BibitemOpen
  \bibfield  {author} {\bibinfo {author} {\bibfnamefont {T.~D.}\ \bibnamefont
  {Phan}}, \bibinfo {author} {\bibfnamefont {J.~P.}\ \bibnamefont {Eastwood}},
  \bibinfo {author} {\bibfnamefont {M.~A.}\ \bibnamefont {Shay}}, \bibinfo
  {author} {\bibfnamefont {J.~F.}\ \bibnamefont {Drake}}, \bibinfo {author}
  {\bibfnamefont {B.~U.}\ \bibnamefont {Sonnerup}}, \bibinfo {author}
  {\bibfnamefont {M.}~\bibnamefont {Fujimoto}}, \bibinfo {author}
  {\bibfnamefont {P.~A.}\ \bibnamefont {Cassak}}, \bibinfo {author}
  {\bibfnamefont {M.}~\bibnamefont {{\O}ieroset}}, \bibinfo {author}
  {\bibfnamefont {J.~L.}\ \bibnamefont {Burch}}, \bibinfo {author}
  {\bibfnamefont {R.~B.}\ \bibnamefont {Torbert}}, \bibinfo {author}
  {\bibfnamefont {A.~C.}\ \bibnamefont {Rager}}, \bibinfo {author}
  {\bibfnamefont {J.~C.}\ \bibnamefont {Dorelli}}, \bibinfo {author}
  {\bibfnamefont {D.~J.}\ \bibnamefont {Gershman}}, \bibinfo {author}
  {\bibfnamefont {C.}~\bibnamefont {Pollock}}, \bibinfo {author} {\bibfnamefont
  {P.~S.}\ \bibnamefont {Pyakurel}}, \bibinfo {author} {\bibfnamefont {C.~C.}\
  \bibnamefont {Haggerty}}, \bibinfo {author} {\bibfnamefont {Y.}~\bibnamefont
  {Khotyaintsev}}, \bibinfo {author} {\bibfnamefont {B.}~\bibnamefont
  {Lavraud}}, \bibinfo {author} {\bibfnamefont {Y.}~\bibnamefont {Saito}},
  \bibinfo {author} {\bibfnamefont {M.}~\bibnamefont {Oka}}, \bibinfo {author}
  {\bibfnamefont {R.~E.}\ \bibnamefont {Ergun}}, \bibinfo {author}
  {\bibfnamefont {A.}~\bibnamefont {Retino}}, \bibinfo {author} {\bibfnamefont
  {O.}~\bibnamefont {Le~Contel}}, \bibinfo {author} {\bibfnamefont {M.~R.}\
  \bibnamefont {Argall}}, \bibinfo {author} {\bibfnamefont {B.~L.}\
  \bibnamefont {Giles}}, \bibinfo {author} {\bibfnamefont {T.~E.}\ \bibnamefont
  {Moore}}, \bibinfo {author} {\bibfnamefont {F.~D.}\ \bibnamefont {Wilder}},
  \bibinfo {author} {\bibfnamefont {R.~J.}\ \bibnamefont {Strangeway}},
  \bibinfo {author} {\bibfnamefont {C.~T.}\ \bibnamefont {Russell}}, \bibinfo
  {author} {\bibfnamefont {P.~A.}\ \bibnamefont {Lindqvist}},\ and\ \bibinfo
  {author} {\bibfnamefont {W.}~\bibnamefont {Magnes}},\ }\bibfield  {title}
  {\bibinfo {title} {{Electron magnetic reconnection without ion coupling in
  Earth's turbulent magnetosheath}},\ }\href
  {https://doi.org/10.1038/s41586-018-0091-5} {\bibfield  {journal} {\bibinfo
  {journal} {Nature}\ }\textbf {\bibinfo {volume} {557}},\ \bibinfo {pages}
  {202} (\bibinfo {year} {2018})}\BibitemShut {NoStop}%
\bibitem [{\citenamefont {Vega}\ \emph {et~al.}(2020)\citenamefont {Vega},
  \citenamefont {Roytershteyn}, \citenamefont {Delzanno},\ and\ \citenamefont
  {Boldyrev}}]{Vega2020Electron-onlyTurbulence}%
  \BibitemOpen
  \bibfield  {author} {\bibinfo {author} {\bibfnamefont {C.}~\bibnamefont
  {Vega}}, \bibinfo {author} {\bibfnamefont {V.}~\bibnamefont {Roytershteyn}},
  \bibinfo {author} {\bibfnamefont {G.~L.}\ \bibnamefont {Delzanno}},\ and\
  \bibinfo {author} {\bibfnamefont {S.}~\bibnamefont {Boldyrev}},\ }\bibfield
  {title} {\bibinfo {title} {{Electron-only Reconnection in
  Kinetic-Alfv{\'{e}}n Turbulence}},\ }\href
  {https://doi.org/10.3847/2041-8213/ab7eba} {\bibfield  {journal} {\bibinfo
  {journal} {The Astrophysical Journal}\ }\textbf {\bibinfo {volume} {893}},\
  \bibinfo {pages} {L10} (\bibinfo {year} {2020})}\BibitemShut {NoStop}%
\bibitem [{\citenamefont {Stawarz}\ \emph {et~al.}(2019)\citenamefont
  {Stawarz}, \citenamefont {Eastwood}, \citenamefont {Phan}, \citenamefont
  {Gingell}, \citenamefont {Shay}, \citenamefont {Burch}, \citenamefont
  {Ergun}, \citenamefont {Giles}, \citenamefont {Gershman}, \citenamefont
  {Contel}, \citenamefont {Lindqvist}, \citenamefont {Russell}, \citenamefont
  {Strangeway}, \citenamefont {Torbert}, \citenamefont {Argall}, \citenamefont
  {Fischer}, \citenamefont {Magnes},\ and\ \citenamefont
  {Franci}}]{Stawarz2019PropertiesMagnetosheath}%
  \BibitemOpen
  \bibfield  {author} {\bibinfo {author} {\bibfnamefont {J.~E.}\ \bibnamefont
  {Stawarz}}, \bibinfo {author} {\bibfnamefont {J.~P.}\ \bibnamefont
  {Eastwood}}, \bibinfo {author} {\bibfnamefont {T.~D.}\ \bibnamefont {Phan}},
  \bibinfo {author} {\bibfnamefont {I.~L.}\ \bibnamefont {Gingell}}, \bibinfo
  {author} {\bibfnamefont {M.~A.}\ \bibnamefont {Shay}}, \bibinfo {author}
  {\bibfnamefont {J.~L.}\ \bibnamefont {Burch}}, \bibinfo {author}
  {\bibfnamefont {R.~E.}\ \bibnamefont {Ergun}}, \bibinfo {author}
  {\bibfnamefont {B.~L.}\ \bibnamefont {Giles}}, \bibinfo {author}
  {\bibfnamefont {D.~J.}\ \bibnamefont {Gershman}}, \bibinfo {author}
  {\bibfnamefont {O.~L.}\ \bibnamefont {Contel}}, \bibinfo {author}
  {\bibfnamefont {P.-A.}\ \bibnamefont {Lindqvist}}, \bibinfo {author}
  {\bibfnamefont {C.~T.}\ \bibnamefont {Russell}}, \bibinfo {author}
  {\bibfnamefont {R.~J.}\ \bibnamefont {Strangeway}}, \bibinfo {author}
  {\bibfnamefont {R.~B.}\ \bibnamefont {Torbert}}, \bibinfo {author}
  {\bibfnamefont {M.~R.}\ \bibnamefont {Argall}}, \bibinfo {author}
  {\bibfnamefont {D.}~\bibnamefont {Fischer}}, \bibinfo {author} {\bibfnamefont
  {W.}~\bibnamefont {Magnes}},\ and\ \bibinfo {author} {\bibfnamefont
  {L.}~\bibnamefont {Franci}},\ }\bibfield  {title} {\bibinfo {title}
  {{Properties of the Turbulence Associated with Electron-only Magnetic
  Reconnection in Earth’s Magnetosheath}},\ }\href
  {https://doi.org/10.3847/2041-8213/ab21c8} {\bibfield  {journal} {\bibinfo
  {journal} {The Astrophysical Journal}\ }\textbf {\bibinfo {volume} {877}},\
  \bibinfo {pages} {L37} (\bibinfo {year} {2019})}\BibitemShut {NoStop}%
\bibitem [{\citenamefont {Sharma~Pyakurel}\ \emph {et~al.}(2019)\citenamefont
  {Sharma~Pyakurel}, \citenamefont {Shay}, \citenamefont {Phan}, \citenamefont
  {Matthaeus}, \citenamefont {Drake}, \citenamefont {TenBarge}, \citenamefont
  {Haggerty}, \citenamefont {Klein}, \citenamefont {Cassak}, \citenamefont
  {Parashar}, \citenamefont {Swisdak},\ and\ \citenamefont
  {Chasapis}}]{SharmaPyakurel2019TransitionTurbulence}%
  \BibitemOpen
  \bibfield  {author} {\bibinfo {author} {\bibfnamefont {P.}~\bibnamefont
  {Sharma~Pyakurel}}, \bibinfo {author} {\bibfnamefont {M.~A.}\ \bibnamefont
  {Shay}}, \bibinfo {author} {\bibfnamefont {T.~D.}\ \bibnamefont {Phan}},
  \bibinfo {author} {\bibfnamefont {W.~H.}\ \bibnamefont {Matthaeus}}, \bibinfo
  {author} {\bibfnamefont {J.~F.}\ \bibnamefont {Drake}}, \bibinfo {author}
  {\bibfnamefont {J.~M.}\ \bibnamefont {TenBarge}}, \bibinfo {author}
  {\bibfnamefont {C.~C.}\ \bibnamefont {Haggerty}}, \bibinfo {author}
  {\bibfnamefont {K.~G.}\ \bibnamefont {Klein}}, \bibinfo {author}
  {\bibfnamefont {P.~A.}\ \bibnamefont {Cassak}}, \bibinfo {author}
  {\bibfnamefont {T.~N.}\ \bibnamefont {Parashar}}, \bibinfo {author}
  {\bibfnamefont {M.}~\bibnamefont {Swisdak}},\ and\ \bibinfo {author}
  {\bibfnamefont {A.}~\bibnamefont {Chasapis}},\ }\bibfield  {title} {\bibinfo
  {title} {{Transition from ion-coupled to electron-only reconnection: Basic
  physics and implications for plasma turbulence}},\ }\href
  {https://doi.org/10.1063/1.5090403} {\bibfield  {journal} {\bibinfo
  {journal} {Physics of Plasmas}\ }\textbf {\bibinfo {volume} {26}},\ \bibinfo
  {pages} {82307} (\bibinfo {year} {2019})}\BibitemShut {NoStop}%
\bibitem [{\citenamefont {Kozak}\ \emph {et~al.}(2011)\citenamefont {Kozak},
  \citenamefont {Pilipenko}, \citenamefont {Chugunova},\ and\ \citenamefont
  {Kozak}}]{Kozak2011StatisticalMagnetosheath}%
  \BibitemOpen
  \bibfield  {author} {\bibinfo {author} {\bibfnamefont {L.~V.}\ \bibnamefont
  {Kozak}}, \bibinfo {author} {\bibfnamefont {V.~A.}\ \bibnamefont
  {Pilipenko}}, \bibinfo {author} {\bibfnamefont {O.~M.}\ \bibnamefont
  {Chugunova}},\ and\ \bibinfo {author} {\bibfnamefont {P.~N.}\ \bibnamefont
  {Kozak}},\ }\bibfield  {title} {\bibinfo {title} {{Statistical analysis of
  turbulence in the foreshock region and in the Earth's magnetosheath}},\
  }\href {https://doi.org/10.1134/S0010952511030063} {\bibfield  {journal}
  {\bibinfo  {journal} {Cosmic Research}\ }\textbf {\bibinfo {volume} {49}},\
  \bibinfo {pages} {194} (\bibinfo {year} {2011})}\BibitemShut {NoStop}%
\bibitem [{\citenamefont {Wang}\ \emph {et~al.}(2012)\citenamefont {Wang},
  \citenamefont {Gkioulidou}, \citenamefont {Lyons},\ and\ \citenamefont
  {Angelopoulos}}]{Wang2012SpatialSheet}%
  \BibitemOpen
  \bibfield  {author} {\bibinfo {author} {\bibfnamefont {C.~P.}\ \bibnamefont
  {Wang}}, \bibinfo {author} {\bibfnamefont {M.}~\bibnamefont {Gkioulidou}},
  \bibinfo {author} {\bibfnamefont {L.~R.}\ \bibnamefont {Lyons}},\ and\
  \bibinfo {author} {\bibfnamefont {V.}~\bibnamefont {Angelopoulos}},\
  }\bibfield  {title} {\bibinfo {title} {{Spatial distributions of the ion to
  electron temperature ratio in the magnetosheath and plasma sheet}},\ }\href
  {https://doi.org/10.1029/2012JA017658} {\bibfield  {journal} {\bibinfo
  {journal} {Journal of Geophysical Research: Space Physics}\ }\textbf
  {\bibinfo {volume} {117}},\ \bibinfo {pages} {1} (\bibinfo {year}
  {2012})}\BibitemShut {NoStop}%
\bibitem [{\citenamefont {Iwai}\ \emph {et~al.}(2014)\citenamefont {Iwai},
  \citenamefont {Shibasaki}, \citenamefont {Nozawa}, \citenamefont {Takahashi},
  \citenamefont {Sawada}, \citenamefont {Kitagawa}, \citenamefont {Miyawaki},\
  and\ \citenamefont {Kashiwagi}}]{Iwai2014CoronalObservations}%
  \BibitemOpen
  \bibfield  {author} {\bibinfo {author} {\bibfnamefont {K.}~\bibnamefont
  {Iwai}}, \bibinfo {author} {\bibfnamefont {K.}~\bibnamefont {Shibasaki}},
  \bibinfo {author} {\bibfnamefont {S.}~\bibnamefont {Nozawa}}, \bibinfo
  {author} {\bibfnamefont {T.}~\bibnamefont {Takahashi}}, \bibinfo {author}
  {\bibfnamefont {S.}~\bibnamefont {Sawada}}, \bibinfo {author} {\bibfnamefont
  {J.}~\bibnamefont {Kitagawa}}, \bibinfo {author} {\bibfnamefont
  {S.}~\bibnamefont {Miyawaki}},\ and\ \bibinfo {author} {\bibfnamefont
  {H.}~\bibnamefont {Kashiwagi}},\ }\bibfield  {title} {\bibinfo {title}
  {{Coronal magnetic field and the plasma beta determined from radio and
  multiple satellite observations}},\ }\href
  {https://doi.org/10.1186/s40623-014-0149-z} {\bibfield  {journal} {\bibinfo
  {journal} {Earth, Planets and Space}\ }\textbf {\bibinfo {volume} {66}},\
  \bibinfo {pages} {1} (\bibinfo {year} {2014})}\BibitemShut {NoStop}%
\bibitem [{\citenamefont {Lucek}\ \emph {et~al.}(2005)\citenamefont {Lucek},
  \citenamefont {Constantinescu}, \citenamefont {Goldstein}, \citenamefont
  {Pickett}, \citenamefont {Pin{\c{c}}on}, \citenamefont {Sahraoui},
  \citenamefont {Treumann},\ and\ \citenamefont
  {Walker}}]{Lucek2005TheMagnetosheath}%
  \BibitemOpen
  \bibfield  {author} {\bibinfo {author} {\bibfnamefont {E.~A.}\ \bibnamefont
  {Lucek}}, \bibinfo {author} {\bibfnamefont {D.}~\bibnamefont
  {Constantinescu}}, \bibinfo {author} {\bibfnamefont {M.~L.}\ \bibnamefont
  {Goldstein}}, \bibinfo {author} {\bibfnamefont {J.}~\bibnamefont {Pickett}},
  \bibinfo {author} {\bibfnamefont {J.~L.}\ \bibnamefont {Pin{\c{c}}on}},
  \bibinfo {author} {\bibfnamefont {F.}~\bibnamefont {Sahraoui}}, \bibinfo
  {author} {\bibfnamefont {R.~A.}\ \bibnamefont {Treumann}},\ and\ \bibinfo
  {author} {\bibfnamefont {S.~N.}\ \bibnamefont {Walker}},\ }\bibfield  {title}
  {\bibinfo {title} {{The magnetosheath}},\ }\href
  {https://doi.org/10.1007/s11214-005-3825-2} {\bibfield  {journal} {\bibinfo
  {journal} {Space Science Reviews}\ }\textbf {\bibinfo {volume} {118}},\
  \bibinfo {pages} {95} (\bibinfo {year} {2005})}\BibitemShut {NoStop}%
\bibitem [{\citenamefont {Sahraoui}\ \emph {et~al.}(2013)\citenamefont
  {Sahraoui}, \citenamefont {Huang}, \citenamefont {Belmont}, \citenamefont
  {Goldstein}, \citenamefont {R{\'{e}}tino}, \citenamefont {Robert},\ and\
  \citenamefont {De~Patoul}}]{Sahraoui2013ScalingTurbulence}%
  \BibitemOpen
  \bibfield  {author} {\bibinfo {author} {\bibfnamefont {F.}~\bibnamefont
  {Sahraoui}}, \bibinfo {author} {\bibfnamefont {S.~Y.}\ \bibnamefont {Huang}},
  \bibinfo {author} {\bibfnamefont {G.}~\bibnamefont {Belmont}}, \bibinfo
  {author} {\bibfnamefont {M.~L.}\ \bibnamefont {Goldstein}}, \bibinfo {author}
  {\bibfnamefont {A.}~\bibnamefont {R{\'{e}}tino}}, \bibinfo {author}
  {\bibfnamefont {P.}~\bibnamefont {Robert}},\ and\ \bibinfo {author}
  {\bibfnamefont {J.}~\bibnamefont {De~Patoul}},\ }\bibfield  {title} {\bibinfo
  {title} {{Scaling of the electron dissipation range of solar wind
  turbulence}},\ }\href {https://doi.org/10.1088/0004-637X/777/1/15} {\bibfield
   {journal} {\bibinfo  {journal} {Astrophysical Journal}\ }\textbf {\bibinfo
  {volume} {777}},\ \bibinfo {pages} {15} (\bibinfo {year} {2013})}\BibitemShut
  {NoStop}%
\bibitem [{\citenamefont {Hughes}\ \emph {et~al.}(2017)\citenamefont {Hughes},
  \citenamefont {Gary},\ and\ \citenamefont
  {Wang}}]{Hughes2017Particle-in-cell}%
  \BibitemOpen
  \bibfield  {author} {\bibinfo {author} {\bibfnamefont {R.~S.}\ \bibnamefont
  {Hughes}}, \bibinfo {author} {\bibfnamefont {S.~P.}\ \bibnamefont {Gary}},\
  and\ \bibinfo {author} {\bibfnamefont {J.}~\bibnamefont {Wang}},\ }\bibfield
  {title} {\bibinfo {title} {{Particle-in-cell Simulations of Electron and Ion
  Dissipation by Whistler Turbulence: Variations with Electron {$\beta$}}},\
  }\href {https://doi.org/10.3847/2041-8213/835/1/l15} {\bibfield  {journal}
  {\bibinfo  {journal} {The Astrophysical Journal}\ }\textbf {\bibinfo {volume}
  {835}},\ \bibinfo {pages} {L15} (\bibinfo {year} {2017})}\BibitemShut
  {NoStop}%
\bibitem [{\citenamefont {Passot}\ \emph {et~al.}(2018)\citenamefont {Passot},
  \citenamefont {Sulem},\ and\ \citenamefont
  {Tassi}}]{Passot2018GyrofluidTurbulence}%
  \BibitemOpen
  \bibfield  {author} {\bibinfo {author} {\bibfnamefont {T.}~\bibnamefont
  {Passot}}, \bibinfo {author} {\bibfnamefont {P.~L.}\ \bibnamefont {Sulem}},\
  and\ \bibinfo {author} {\bibfnamefont {E.}~\bibnamefont {Tassi}},\ }\bibfield
   {title} {\bibinfo {title} {{Gyrofluid modeling and phenomenology of
  low-{$\beta$}e Alfv{\'{e}}n wave turbulence}},\ }\href
  {https://doi.org/10.1063/1.5022528} {\bibfield  {journal} {\bibinfo
  {journal} {Physics of Plasmas}\ }\textbf {\bibinfo {volume} {25}},\ \bibinfo
  {pages} {42107} (\bibinfo {year} {2018})}\BibitemShut {NoStop}%
\bibitem [{\citenamefont {Matthaeus}\ \emph {et~al.}(2015)\citenamefont
  {Matthaeus}, \citenamefont {Wan}, \citenamefont {Servidio}, \citenamefont
  {Greco}, \citenamefont {Osman}, \citenamefont {Oughton},\ and\ \citenamefont
  {Dmitruk}}]{Matthaeus2015IntermittencyPlasmas}%
  \BibitemOpen
  \bibfield  {author} {\bibinfo {author} {\bibfnamefont {W.~H.}\ \bibnamefont
  {Matthaeus}}, \bibinfo {author} {\bibfnamefont {M.}~\bibnamefont {Wan}},
  \bibinfo {author} {\bibfnamefont {S.}~\bibnamefont {Servidio}}, \bibinfo
  {author} {\bibfnamefont {A.}~\bibnamefont {Greco}}, \bibinfo {author}
  {\bibfnamefont {K.~T.}\ \bibnamefont {Osman}}, \bibinfo {author}
  {\bibfnamefont {S.}~\bibnamefont {Oughton}},\ and\ \bibinfo {author}
  {\bibfnamefont {P.}~\bibnamefont {Dmitruk}},\ }\bibfield  {title} {\bibinfo
  {title} {{Intermittency, nonlinear dynamics and dissipation in the solar wind
  and astrophysical plasmas}},\ }\bibfield  {journal} {\bibinfo  {journal}
  {Philosophical Transactions of the Royal Society A: Mathematical, Physical
  and Engineering Sciences}\ }\textbf {\bibinfo {volume} {373}},\ \href
  {https://doi.org/10.1098/rsta.2014.0154} {10.1098/rsta.2014.0154} (\bibinfo
  {year} {2015})\BibitemShut {NoStop}%
\bibitem [{\citenamefont {Alexakis}\ and\ \citenamefont
  {Biferale}(2018)}]{Alexakis2018CascadesFlows}%
  \BibitemOpen
  \bibfield  {author} {\bibinfo {author} {\bibfnamefont {A.}~\bibnamefont
  {Alexakis}}\ and\ \bibinfo {author} {\bibfnamefont {L.}~\bibnamefont
  {Biferale}},\ }\bibfield  {title} {\bibinfo {title} {{Cascades and
  transitions in turbulent flows}},\ }\href
  {https://doi.org/10.1016/j.physrep.2018.08.001} {\bibfield  {journal}
  {\bibinfo  {journal} {Physics Reports}\ }\textbf {\bibinfo {volume}
  {767-769}},\ \bibinfo {pages} {1} (\bibinfo {year} {2018})}\BibitemShut
  {NoStop}%
\bibitem [{\citenamefont {Brissaud}\ \emph {et~al.}(1973)\citenamefont
  {Brissaud}, \citenamefont {Frisch}, \citenamefont {Leorat}, \citenamefont
  {Lesieur},\ and\ \citenamefont {Mazure}}]{Brissaud1973HelicityTurbulence}%
  \BibitemOpen
  \bibfield  {author} {\bibinfo {author} {\bibfnamefont {A.}~\bibnamefont
  {Brissaud}}, \bibinfo {author} {\bibfnamefont {U.}~\bibnamefont {Frisch}},
  \bibinfo {author} {\bibfnamefont {J.}~\bibnamefont {Leorat}}, \bibinfo
  {author} {\bibfnamefont {M.}~\bibnamefont {Lesieur}},\ and\ \bibinfo {author}
  {\bibfnamefont {A.}~\bibnamefont {Mazure}},\ }\bibfield  {title} {\bibinfo
  {title} {{Helicity cascades in fully developed isotropic turbulence}},\
  }\href {https://doi.org/10.1063/1.1694520} {\bibfield  {journal} {\bibinfo
  {journal} {Physics of Fluids}\ }\textbf {\bibinfo {volume} {16}},\ \bibinfo
  {pages} {1366} (\bibinfo {year} {1973})}\BibitemShut {NoStop}%
\bibitem [{\citenamefont {Chen}\ \emph {et~al.}(2003)\citenamefont {Chen},
  \citenamefont {Chen}, \citenamefont {Eyink},\ and\ \citenamefont
  {Holm}}]{Chen2003IntermittencyHelicity}%
  \BibitemOpen
  \bibfield  {author} {\bibinfo {author} {\bibfnamefont {Q.}~\bibnamefont
  {Chen}}, \bibinfo {author} {\bibfnamefont {S.}~\bibnamefont {Chen}}, \bibinfo
  {author} {\bibfnamefont {G.~L.}\ \bibnamefont {Eyink}},\ and\ \bibinfo
  {author} {\bibfnamefont {D.~D.}\ \bibnamefont {Holm}},\ }\bibfield  {title}
  {\bibinfo {title} {{Intermittency in the Joint Cascade of Energy and
  Helicity}},\ }\href {https://doi.org/10.1103/PhysRevLett.90.214503}
  {\bibfield  {journal} {\bibinfo  {journal} {Physical Review Letters}\
  }\textbf {\bibinfo {volume} {90}},\ \bibinfo {pages} {4} (\bibinfo {year}
  {2003})}\BibitemShut {NoStop}%
\bibitem [{\citenamefont {Podesta}\ and\ \citenamefont
  {Bhattacharjee}(2010)}]{Podesta2010THEORYCROSS-HELICITY}%
  \BibitemOpen
  \bibfield  {author} {\bibinfo {author} {\bibfnamefont {J.~J.}\ \bibnamefont
  {Podesta}}\ and\ \bibinfo {author} {\bibfnamefont {A.}~\bibnamefont
  {Bhattacharjee}},\ }\bibfield  {title} {\bibinfo {title} {{Theory of
  incompressible magnetohydrodynamic turbulence with scale-dependent alignment
  and cross-helicity}},\ }\href {https://doi.org/10.1088/0004-637X/718/2/1151}
  {\bibfield  {journal} {\bibinfo  {journal} {Astrophysical Journal}\ }\textbf
  {\bibinfo {volume} {718}},\ \bibinfo {pages} {1151} (\bibinfo {year}
  {2010})}\BibitemShut {NoStop}%
\bibitem [{\citenamefont {Sahoo}\ \emph {et~al.}(2017)\citenamefont {Sahoo},
  \citenamefont {De~Pietro},\ and\ \citenamefont
  {Biferale}}]{Sahoo2017HelicityModels}%
  \BibitemOpen
  \bibfield  {author} {\bibinfo {author} {\bibfnamefont {G.}~\bibnamefont
  {Sahoo}}, \bibinfo {author} {\bibfnamefont {M.}~\bibnamefont {De~Pietro}},\
  and\ \bibinfo {author} {\bibfnamefont {L.}~\bibnamefont {Biferale}},\
  }\bibfield  {title} {\bibinfo {title} {{Helicity statistics in homogeneous
  and isotropic turbulence and turbulence models}},\ }\href
  {https://doi.org/10.1103/PhysRevFluids.2.024601} {\bibfield  {journal}
  {\bibinfo  {journal} {Physical Review Fluids}\ }\textbf {\bibinfo {volume}
  {2}},\ \bibinfo {pages} {24601} (\bibinfo {year} {2017})}\BibitemShut
  {NoStop}%
\bibitem [{\citenamefont {Pouquet}\ \emph {et~al.}(2019)\citenamefont
  {Pouquet}, \citenamefont {Rosenberg}, \citenamefont {Stawarz},\ and\
  \citenamefont {Marino}}]{Pouquet2019HelicityReview}%
  \BibitemOpen
  \bibfield  {author} {\bibinfo {author} {\bibfnamefont {A.}~\bibnamefont
  {Pouquet}}, \bibinfo {author} {\bibfnamefont {D.}~\bibnamefont {Rosenberg}},
  \bibinfo {author} {\bibfnamefont {J.~E.}\ \bibnamefont {Stawarz}},\ and\
  \bibinfo {author} {\bibfnamefont {R.}~\bibnamefont {Marino}},\ }\bibfield
  {title} {\bibinfo {title} {{Helicity Dynamics, Inverse, and Bidirectional
  Cascades in Fluid and Magnetohydrodynamic Turbulence: A Brief Review}},\
  }\href {https://doi.org/10.1029/2018EA000432} {\bibfield  {journal} {\bibinfo
   {journal} {Earth and Space Science}\ }\textbf {\bibinfo {volume} {6}},\
  \bibinfo {pages} {351} (\bibinfo {year} {2019})}\BibitemShut {NoStop}%
\bibitem [{Note1()}]{Note1}%
  \BibitemOpen
  \bibinfo {note} {The dispersion relation for inertial Alfv\'en waves, Eq.
  (\ref {eq:eigenfunction}), implies that $\omega _l/|k_z| \gg v_{the}$ at
  scales $k_{\perp } \rho _e \ll 1$, where $\rho _e = m_e c v_{the}/e B_0$ is
  the electron Larmor radius. Since low-$\beta _e$ implies $d_e \gg \rho _e$,
  we find that electron Landau damping may thus be neglected at scales $d_e >
  k_{\perp }^{-1} \gg \rho _e$, where the resonant condition is not satisfied.
  This is unlike $\beta _e\sim 1$ plasmas, in which $d_e \sim \rho _e$ and
  where, therefore, electron Landau damping may be expected to play an
  important role at electron scales \cite
  {Sahraoui2009EvidenceGyroscale,Alexandrova2013SolarInstabilities}.}\BibitemShut
  {Stop}%
\bibitem [{\citenamefont {Loureiro}\ and\ \citenamefont
  {Boldyrev}(2018)}]{Loureiro2018TurbulencePlasmasc}%
  \BibitemOpen
  \bibfield  {author} {\bibinfo {author} {\bibfnamefont {N.~F.}\ \bibnamefont
  {Loureiro}}\ and\ \bibinfo {author} {\bibfnamefont {S.}~\bibnamefont
  {Boldyrev}},\ }\bibfield  {title} {\bibinfo {title} {{Turbulence in
  Magnetized Pair Plasmas}},\ }\href {https://doi.org/10.3847/2041-8213/aae483}
  {\bibfield  {journal} {\bibinfo  {journal} {The Astrophysical Journal}\
  }\textbf {\bibinfo {volume} {866}},\ \bibinfo {pages} {L14} (\bibinfo {year}
  {2018})}\BibitemShut {NoStop}%
\bibitem [{\citenamefont {Zocco}\ and\ \citenamefont
  {Schekochihin}(2011)}]{Zocco2011}%
  \BibitemOpen
  \bibfield  {author} {\bibinfo {author} {\bibfnamefont {A.}~\bibnamefont
  {Zocco}}\ and\ \bibinfo {author} {\bibfnamefont {A.~A.}\ \bibnamefont
  {Schekochihin}},\ }\bibfield  {title} {\bibinfo {title} {{Reduced
  fluid-kinetic equations for low-frequency dynamics, magnetic reconnection,
  and electron heating in low-beta plasmas}},\ }\href
  {https://doi.org/10.1063/1.3628639} {\bibfield  {journal} {\bibinfo
  {journal} {Physics of Plasmas}\ }\textbf {\bibinfo {volume} {18}},\ \bibinfo
  {pages} {1} (\bibinfo {year} {2011})}\BibitemShut {NoStop}%
\bibitem [{\citenamefont {Boldyrev}\ and\ \citenamefont
  {Loureiro}(2019)}]{Boldyrev2019RoleTurbulence}%
  \BibitemOpen
  \bibfield  {author} {\bibinfo {author} {\bibfnamefont {S.}~\bibnamefont
  {Boldyrev}}\ and\ \bibinfo {author} {\bibfnamefont {N.~F.}\ \bibnamefont
  {Loureiro}},\ }\bibfield  {title} {\bibinfo {title} {{Role of reconnection in
  inertial kinetic-Alfv{\'{e}}n turbulence}},\ }\bibfield  {journal} {\bibinfo
  {journal} {Physical Review Research}\ }\textbf {\bibinfo {volume} {1}},\
  \href {https://doi.org/10.1103/physrevresearch.1.012006}
  {10.1103/physrevresearch.1.012006} (\bibinfo {year} {2019})\BibitemShut
  {NoStop}%
\bibitem [{\citenamefont {Julien}\ and\ \citenamefont
  {Knobloch}(2007)}]{Julien2007ReducedConstraints}%
  \BibitemOpen
  \bibfield  {author} {\bibinfo {author} {\bibfnamefont {K.}~\bibnamefont
  {Julien}}\ and\ \bibinfo {author} {\bibfnamefont {E.}~\bibnamefont
  {Knobloch}},\ }\bibfield  {title} {\bibinfo {title} {{Reduced models for
  fluid flows with strong constraints}},\ }\href
  {https://doi.org/10.1063/1.2741042} {\bibfield  {journal} {\bibinfo
  {journal} {Journal of Mathematical Physics}\ }\textbf {\bibinfo {volume}
  {48}},\ \bibinfo {pages} {65405} (\bibinfo {year} {2007})}\BibitemShut
  {NoStop}%
\bibitem [{\citenamefont {Kadomtsev}\ and\ \citenamefont
  {Pogutse}(1974)}]{Kadomtsev1974NonlinearTokamak}%
  \BibitemOpen
  \bibfield  {author} {\bibinfo {author} {\bibfnamefont {B.}~\bibnamefont
  {Kadomtsev}}\ and\ \bibinfo {author} {\bibfnamefont {O.}~\bibnamefont
  {Pogutse}},\ }\bibfield  {title} {\bibinfo {title} {{Nonlinear helical
  perturbations of a plasma in the tokamak}},\ }\href@noop {} {\bibfield
  {journal} {\bibinfo  {journal} {Soviet Journal of Experimental and
  Theoretical Physics}\ }\textbf {\bibinfo {volume} {38}},\ \bibinfo {pages}
  {575} (\bibinfo {year} {1974})}\BibitemShut {NoStop}%
\bibitem [{\citenamefont {Strauss}(1976)}]{Strauss1976}%
  \BibitemOpen
  \bibfield  {author} {\bibinfo {author} {\bibfnamefont {H.~R.}\ \bibnamefont
  {Strauss}},\ }\bibfield  {title} {\bibinfo {title} {{Nonlinear,
  three-dimensional magnetohydrodynamics of noncircular tokamaks}},\ }\bibfield
   {journal} {\bibinfo  {journal} {The Physics of Fluids}\ }\textbf {\bibinfo
  {volume} {19}},\ \href {https://doi.org/10.1063/1.861310} {10.1063/1.861310}
  (\bibinfo {year} {1976})\BibitemShut {NoStop}%
\bibitem [{\citenamefont {Schekochihin}\ \emph {et~al.}(2009)\citenamefont
  {Schekochihin}, \citenamefont {Cowley}, \citenamefont {Dorland},
  \citenamefont {Hammett}, \citenamefont {Howes}, \citenamefont {Quataert},\
  and\ \citenamefont {Tatsuno}}]{Schekochihin2009AstrophysicalPlasmas}%
  \BibitemOpen
  \bibfield  {author} {\bibinfo {author} {\bibfnamefont {A.~A.}\ \bibnamefont
  {Schekochihin}}, \bibinfo {author} {\bibfnamefont {S.~C.}\ \bibnamefont
  {Cowley}}, \bibinfo {author} {\bibfnamefont {W.}~\bibnamefont {Dorland}},
  \bibinfo {author} {\bibfnamefont {G.~W.}\ \bibnamefont {Hammett}}, \bibinfo
  {author} {\bibfnamefont {G.~G.}\ \bibnamefont {Howes}}, \bibinfo {author}
  {\bibfnamefont {E.}~\bibnamefont {Quataert}},\ and\ \bibinfo {author}
  {\bibfnamefont {T.}~\bibnamefont {Tatsuno}},\ }\bibfield  {title} {\bibinfo
  {title} {{Astrophysical gyrokinetics: Kinetic and fluid turbulent cascades in
  magnetized weakly collisional plasmas}},\ }\href
  {https://doi.org/10.1088/0067-0049/182/1/310} {\bibfield  {journal} {\bibinfo
   {journal} {Astrophysical Journal, Supplement Series}\ }\textbf {\bibinfo
  {volume} {182}},\ \bibinfo {pages} {310} (\bibinfo {year}
  {2009})}\BibitemShut {NoStop}%
\bibitem [{\citenamefont {Perez}\ \emph {et~al.}(2012)\citenamefont {Perez},
  \citenamefont {Mason}, \citenamefont {Boldyrev},\ and\ \citenamefont
  {Cattaneo}}]{Perez2012OnTurbulence}%
  \BibitemOpen
  \bibfield  {author} {\bibinfo {author} {\bibfnamefont {J.~C.}\ \bibnamefont
  {Perez}}, \bibinfo {author} {\bibfnamefont {J.}~\bibnamefont {Mason}},
  \bibinfo {author} {\bibfnamefont {S.}~\bibnamefont {Boldyrev}},\ and\
  \bibinfo {author} {\bibfnamefont {F.}~\bibnamefont {Cattaneo}},\ }\bibfield
  {title} {\bibinfo {title} {{On the energy spectrum of strong
  magnetohydrodynamic turbulence}},\ }\bibfield  {journal} {\bibinfo  {journal}
  {Physical Review X}\ }\textbf {\bibinfo {volume} {2}},\ \href
  {https://doi.org/10.1103/PhysRevX.2.041005} {10.1103/PhysRevX.2.041005}
  (\bibinfo {year} {2012})\BibitemShut {NoStop}%
\bibitem [{\citenamefont {Goldreich}\ and\ \citenamefont
  {Sridhar}(1995)}]{Goldreich1995TowardTurbulence}%
  \BibitemOpen
  \bibfield  {author} {\bibinfo {author} {\bibfnamefont {P.}~\bibnamefont
  {Goldreich}}\ and\ \bibinfo {author} {\bibfnamefont {S.}~\bibnamefont
  {Sridhar}},\ }\bibfield  {title} {\bibinfo {title} {{Toward a theory of
  interstellar turbulence. 2: Strong alfvenic turbulence}},\ }\href
  {https://doi.org/10.1086/175121} {\bibfield  {journal} {\bibinfo  {journal}
  {The Astrophysical Journal}\ }\textbf {\bibinfo {volume} {438}},\ \bibinfo
  {pages} {763} (\bibinfo {year} {1995})}\BibitemShut {NoStop}%
\bibitem [{\citenamefont {Loureiro}\ \emph {et~al.}(2016)\citenamefont
  {Loureiro}, \citenamefont {Dorland}, \citenamefont {Fazendeiro},
  \citenamefont {Kanekar}, \citenamefont {Mallet}, \citenamefont {Vilelas},\
  and\ \citenamefont {Zocco}}]{Loureiro2016}%
  \BibitemOpen
  \bibfield  {author} {\bibinfo {author} {\bibfnamefont {N.~F.}\ \bibnamefont
  {Loureiro}}, \bibinfo {author} {\bibfnamefont {W.}~\bibnamefont {Dorland}},
  \bibinfo {author} {\bibfnamefont {L.}~\bibnamefont {Fazendeiro}}, \bibinfo
  {author} {\bibfnamefont {A.}~\bibnamefont {Kanekar}}, \bibinfo {author}
  {\bibfnamefont {A.}~\bibnamefont {Mallet}}, \bibinfo {author} {\bibfnamefont
  {M.~S.}\ \bibnamefont {Vilelas}},\ and\ \bibinfo {author} {\bibfnamefont
  {A.}~\bibnamefont {Zocco}},\ }\bibfield  {title} {\bibinfo {title} {{Viriato:
  A Fourier–Hermite spectral code for strongly magnetized fluid–kinetic
  plasma dynamics}},\ }\href {https://doi.org/10.1016/j.cpc.2016.05.004}
  {\bibfield  {journal} {\bibinfo  {journal} {Computer Physics Communications}\
  }\textbf {\bibinfo {volume} {206}},\ \bibinfo {pages} {45} (\bibinfo {year}
  {2016})}\BibitemShut {NoStop}%
\bibitem [{\citenamefont {Cho}\ and\ \citenamefont
  {Lazarian}(2004)}]{Cho2004TheTurbulence}%
  \BibitemOpen
  \bibfield  {author} {\bibinfo {author} {\bibfnamefont {J.}~\bibnamefont
  {Cho}}\ and\ \bibinfo {author} {\bibfnamefont {A.}~\bibnamefont {Lazarian}},\
  }\bibfield  {title} {\bibinfo {title} {{The Anisotropy of Electron
  Magnetohydrodynamic Turbulence}},\ }\href {https://doi.org/10.1086/425215}
  {\bibfield  {journal} {\bibinfo  {journal} {The Astrophysical Journal}\
  }\textbf {\bibinfo {volume} {615}},\ \bibinfo {pages} {L41} (\bibinfo {year}
  {2004})}\BibitemShut {NoStop}%
\bibitem [{\citenamefont {Stenson}\ \emph {et~al.}(2018)\citenamefont
  {Stenson}, \citenamefont {Ni{\ss}l}, \citenamefont {Hergenhahn},
  \citenamefont {Horn-Stanja}, \citenamefont {Singer}, \citenamefont {Saitoh},
  \citenamefont {Sunn~Pedersen}, \citenamefont {Danielson}, \citenamefont
  {Stoneking}, \citenamefont {Dickmann},\ and\ \citenamefont
  {Hugenschmidt}}]{Stenson2018LosslessTrap}%
  \BibitemOpen
  \bibfield  {author} {\bibinfo {author} {\bibfnamefont {E.~V.}\ \bibnamefont
  {Stenson}}, \bibinfo {author} {\bibfnamefont {S.}~\bibnamefont {Ni{\ss}l}},
  \bibinfo {author} {\bibfnamefont {U.}~\bibnamefont {Hergenhahn}}, \bibinfo
  {author} {\bibfnamefont {J.}~\bibnamefont {Horn-Stanja}}, \bibinfo {author}
  {\bibfnamefont {M.}~\bibnamefont {Singer}}, \bibinfo {author} {\bibfnamefont
  {H.}~\bibnamefont {Saitoh}}, \bibinfo {author} {\bibfnamefont
  {T.}~\bibnamefont {Sunn~Pedersen}}, \bibinfo {author} {\bibfnamefont {J.~R.}\
  \bibnamefont {Danielson}}, \bibinfo {author} {\bibfnamefont {M.~R.}\
  \bibnamefont {Stoneking}}, \bibinfo {author} {\bibfnamefont {M.}~\bibnamefont
  {Dickmann}},\ and\ \bibinfo {author} {\bibfnamefont {C.}~\bibnamefont
  {Hugenschmidt}},\ }\bibfield  {title} {\bibinfo {title} {{Lossless Positron
  Injection into a Magnetic Dipole Trap}},\ }\bibfield  {journal} {\bibinfo
  {journal} {Physical Review Letters}\ }\textbf {\bibinfo {volume} {121}},\
  \href {https://doi.org/10.1103/PhysRevLett.121.235005}
  {10.1103/PhysRevLett.121.235005} (\bibinfo {year} {2018})\BibitemShut
  {NoStop}%
\bibitem [{\citenamefont {Horn-Stanja}\ \emph {et~al.}(2020)\citenamefont
  {Horn-Stanja}, \citenamefont {Stenson}, \citenamefont {Stoneking},
  \citenamefont {Singer}, \citenamefont {Hergenhahn}, \citenamefont {Ni{\ss}l},
  \citenamefont {Saitoh}, \citenamefont {Pedersen}, \citenamefont {Dickmann},
  \citenamefont {Hugenschmidt},\ and\ \citenamefont
  {Danielson}}]{Horn-Stanja2020InjectionTrapb}%
  \BibitemOpen
  \bibfield  {author} {\bibinfo {author} {\bibfnamefont {J.}~\bibnamefont
  {Horn-Stanja}}, \bibinfo {author} {\bibfnamefont {E.~V.}\ \bibnamefont
  {Stenson}}, \bibinfo {author} {\bibfnamefont {M.~R.}\ \bibnamefont
  {Stoneking}}, \bibinfo {author} {\bibfnamefont {M.}~\bibnamefont {Singer}},
  \bibinfo {author} {\bibfnamefont {U.}~\bibnamefont {Hergenhahn}}, \bibinfo
  {author} {\bibfnamefont {S.}~\bibnamefont {Ni{\ss}l}}, \bibinfo {author}
  {\bibfnamefont {H.}~\bibnamefont {Saitoh}}, \bibinfo {author} {\bibfnamefont
  {S.}~\bibnamefont {Pedersen}}, \bibinfo {author} {\bibfnamefont
  {M.}~\bibnamefont {Dickmann}}, \bibinfo {author} {\bibfnamefont
  {C.}~\bibnamefont {Hugenschmidt}},\ and\ \bibinfo {author} {\bibfnamefont
  {J.~R.}\ \bibnamefont {Danielson}},\ }\bibfield  {title} {\bibinfo {title}
  {{Injection of intense low-energy reactor-based positron beams into a
  supported magnetic dipole trap}},\ }\href
  {https://doi.org/10.1088/2516-1067/ab6f44} {\bibfield  {journal} {\bibinfo
  {journal} {Plasma Res. Express}\ }\textbf {\bibinfo {volume} {2}},\ \bibinfo
  {pages} {15006} (\bibinfo {year} {2020})}\BibitemShut {NoStop}%
\bibitem [{\citenamefont {Nazarenko}\ and\ \citenamefont
  {Schekochihin}(2011)}]{Nazarenko2011CriticalConjecture}%
  \BibitemOpen
  \bibfield  {author} {\bibinfo {author} {\bibfnamefont {S.~V.}\ \bibnamefont
  {Nazarenko}}\ and\ \bibinfo {author} {\bibfnamefont {A.~A.}\ \bibnamefont
  {Schekochihin}},\ }\bibfield  {title} {\bibinfo {title} {{Critical balance in
  magnetohydrodynamic, rotating and stratified turbulence: Towards a universal
  scaling conjecture}},\ }\href {https://doi.org/10.1017/S002211201100067X}
  {\bibfield  {journal} {\bibinfo  {journal} {Journal of Fluid Mechanics}\
  }\textbf {\bibinfo {volume} {677}},\ \bibinfo {pages} {134} (\bibinfo {year}
  {2011})}\BibitemShut {NoStop}%
\bibitem [{\citenamefont {Sahraoui}\ \emph {et~al.}(2009)\citenamefont
  {Sahraoui}, \citenamefont {Goldstein}, \citenamefont {Robert},\ and\
  \citenamefont {Khotyaintsev}}]{Sahraoui2009EvidenceGyroscale}%
  \BibitemOpen
  \bibfield  {author} {\bibinfo {author} {\bibfnamefont {F.}~\bibnamefont
  {Sahraoui}}, \bibinfo {author} {\bibfnamefont {M.~L.}\ \bibnamefont
  {Goldstein}}, \bibinfo {author} {\bibfnamefont {P.}~\bibnamefont {Robert}},\
  and\ \bibinfo {author} {\bibfnamefont {Y.~V.}\ \bibnamefont {Khotyaintsev}},\
  }\bibfield  {title} {\bibinfo {title} {{Evidence of a cascade and dissipation
  of solar-wind turbulence at the electron gyroscale}},\ }\href
  {https://doi.org/10.1103/PhysRevLett.102.231102} {\bibfield  {journal}
  {\bibinfo  {journal} {Physical Review Letters}\ }\textbf {\bibinfo {volume}
  {102}},\ \bibinfo {pages} {1} (\bibinfo {year} {2009})}\BibitemShut {NoStop}%
\bibitem [{\citenamefont {Alexandrova}\ \emph {et~al.}(2013)\citenamefont
  {Alexandrova}, \citenamefont {Chen}, \citenamefont {Sorriso-Valvo},
  \citenamefont {Horbury},\ and\ \citenamefont
  {Bale}}]{Alexandrova2013SolarInstabilities}%
  \BibitemOpen
  \bibfield  {author} {\bibinfo {author} {\bibfnamefont {O.}~\bibnamefont
  {Alexandrova}}, \bibinfo {author} {\bibfnamefont {C.~H.}\ \bibnamefont
  {Chen}}, \bibinfo {author} {\bibfnamefont {L.}~\bibnamefont {Sorriso-Valvo}},
  \bibinfo {author} {\bibfnamefont {T.~S.}\ \bibnamefont {Horbury}},\ and\
  \bibinfo {author} {\bibfnamefont {S.~D.}\ \bibnamefont {Bale}},\ }\bibfield
  {title} {\bibinfo {title} {{Solar wind turbulence and the role of ion
  instabilities}},\ }\href {https://doi.org/10.1007/s11214-013-0004-8}
  {\bibfield  {journal} {\bibinfo  {journal} {Space Science Reviews}\ }\textbf
  {\bibinfo {volume} {178}},\ \bibinfo {pages} {101} (\bibinfo {year}
  {2013})}\BibitemShut {NoStop}%
\end{thebibliography}%


\end{document}